\documentstyle[psfig,epsf,conf-X,10pt]{article}

\def\hawaii{Hawai$'$i~}

\def\arcsec{\ifmmode^{\prime\prime}\;\else$^{\prime\prime}\;$\fi}

\begin{document} 
\small
\heading{A Cluster Deficit in the ROSAT NEP Survey}
\par\medskip\noindent

\author{Isabella M. Gioia$^{1,2}$}
\address{Institute for Astronomy, 2680 Woodlawn Drive, Honolulu, \hawaii~96822, USA}
\address{Istituto di Radioastronomia del CNR, Via Gobetti 101, 40129 Bologna, Italy}

\begin{abstract}
We have used  data from the deepest region of the ROSAT 
All$-$Sky Survey, the North Ecliptic Pole (NEP) region, 
to produce a  complete and unbiased X$-$ray selected sample  of 
distant clusters to understand the nature of cluster evolution 
and determine implications for large scale structure models. 
In this contribution
results are presented from a comparison between the number of 
the observed clusters in the NEP survey and the number of 
expected clusters  assuming no$-$evolution  models. There is a 
deficit by a factor of 2.5$-$4 of high luminosity, high redshift
clusters  with respect to the present. The evolution goes in the  
same direction as the original EMSS  result, and the results from 
the CfA$-$IfA 160 deg$^{2}$ survey by \cite{vik98}. 

\end{abstract}

\section{Introduction}

The ROSAT North Ecliptic Pole (NEP) Survey began nine years
ago right after the completion of the Einstein Medium Sensitivity 
Survey (EMSS) project (\cite{gio90a}). The goal was to construct a 
statistically complete sample of galaxy clusters to

\noindent

1.  study the controversial issue of the cluster X$-$ray Luminosity 
Function (XLF) evolution (\cite{gio90b}; \cite{hen92});

2. characterize the three$-$dimensional large scale structure of 
the universe. 

\noindent
The main difference between the NEP survey and the existing X$-$ray 
serendipitous ROSAT cluster surveys  is that the NEP survey is carried 
out on a contiguous area of sky. Thus our database  will allow 
to examine large scale structure in the cluster distribution. 
In addition, unlike existing X$-$ray cluster surveys, the NEP survey 
is completely optically identified. Specifically, all X$-$rays sources 
in the 81 deg$^{2}$ region have been identified. Throughout
this paper  H$_{0}=$50 km s$^{-1}$ Mpc $^{-1}$, q$_{0}=$0.5 and 
$\Omega_{0}=$1 are used.

\section{The NEP Survey}
Our group at the Institute for Astronomy in \hawaii
has been involved for many years in the optical identification 
of all the sources found in the NEP region  of the  ROSAT
All$-$Sky Survey (RASS, \cite{tru91}; \cite{vog99}).
The NEP region is the deepest area of the RASS where the
ROSAT satellite scan circles overlap and the effective
exposure time exceeds 35ks. The 9$-$year long identification program
has been finally completed.

A total of 446 X$-$ray sources were detected at $>4\sigma$ in the
$0.1-2.4$ keV band using the RASS$-$II processing (described in detail
in \cite{vog99}). We have spectroscopically identified all but three
sources in the survey (see Fig. 1). Redshifts have been measured for the
extragalactic population. We have extracted a complete and
unbiased  sample of 65 galaxy clusters. Twenty clusters have
a redshift greater than 0.3 with the highest being at z$=$0.81.

\begin{figure}
\centerline{\vbox{
\psfig{figure=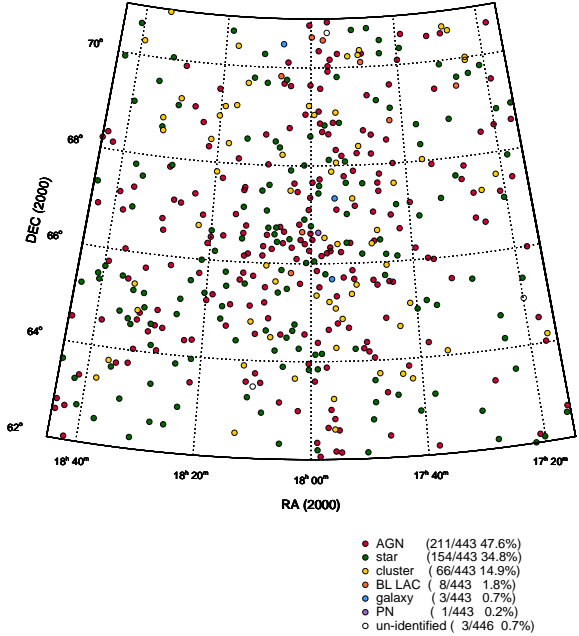,height=6.5cm}
}}
\vskip 2.0truecm
\caption[]{Spatial distribution of all the NEP sources}
\end{figure}

\section{Serendipitous discoveries}

As it happens in the course of identifying a large number of
X$-$ray sources, several serendipitous discoveries were made. Among 
them: the most distant X$-$ray selected QSO at the time of publication
(z$=$4.3, \cite{hen94}); a distant cluster at 0.81 ({\cite{gio99}, 
\cite{hen97}), the second most distant X$-$ray selected cluster so 
far published and the only one with a large number of spectroscopically 
determined cluster member velocities; a supercluster of 20 clusters
at z=0.0877 which is reported in the contribution by
C. Mullis to this meeting. 

\section{A Deficit of High Redshift, High Luminosity Clusters}

The XLF evolution result of \cite{gio90b} inspired many EMSS$-$style
cluster surveys all based on ROSAT archival deep pointing images.
Among them: the RDCS (Rosat Deep Cluster Survey, \cite{ros98});
the WARPS  (Wide Angle Pointed Rosat Survey, \cite{sch97}, \cite{jon98});
the SHARC (Serendipitous High$-$Redshift Archival Rosat Cluster survey,
\cite{col97}, \cite{bur97}, \cite{rom99}, \cite{nic99}) and
finally the largest area survey after the EMSS, the \cite{vik98} 
CfA$-$IfA 160 deg$^{2}$ survey. Each one of these surveys covers an
area of sky of less than 200 deg$^{2}$, much less than the
$\sim800$ deg$^{2}$ of the original EMSS, but with  sensitivities 
almost an order of magnitude deeper than the
EMSS ($\sim 1.8\times10^{-14}$ erg cm$^{-2}$ s$^{-1}$ vs
$\sim 1.3\times10^{-13}$ erg cm$^{-2}$ s$^{-1}$ in 0.3$-$3.5 keV
\footnote{The conversion from 0.5$-$2.0 keV to
0.3$-$3.5 keV is a multiplicative factor of 1.8, assuming a
Raymond$-$Smith with a kT$=$6.0 keV and the standard 0.3
solar abundance.}).
Most of these surveys are still works in progress but a few
preliminary results are available. While all the existing surveys are
in agreement for cluster with the z$<$0.3 and 
$L_{(0.5-2)}<3\times10^{44}$ erg s$^{-1}$, there is not a 
consensus yet  for the brightest and most distant clusters known.

To compare the number of observed clusters with the number of
expected clusters assuming no$-$evolution models, we proceeded as 
follows. In all calculations K$-$corrections have been applied to 
the NEP clusters by assuming for each cluster a temperature derived 
from the L$_{X}-T_{X}$ relation of \cite{whi97}. A constant cluster 
core radius of 0.25 Mpc has been assumed.
\footnote{As shown by \cite{vik98} the distributions of core
radii for nearby (z$<$0.2) and distant (z$>$0.4) clusters are very
similar with a  difference for the average radius at z$>$0.4 by a
factor of only 0.9$\pm$0.1.}  The local (z$<$0.3)
luminosity functions derived from the Southern hemisphere RASS1
Bright Sample  (\cite{deg99}) and from the Northern hemisphere BCS
(\cite{ebe97}) have been integrated in the appropriate  redshift
and luminosity ranges. First the two luminosity functions  are integrated
between z$=$0.3 and z$=$0.85 and between L$_{min}=2\times10^{42}$
and L$_{max}=10^{47}$ erg s$^{-1}$. 
A value of 59.9 (39.9) clusters are expected according to the
RASS1 (BCS), and only 20 NEP clusters are observed. The result is a
factor of 3 (2) less than predicted  from the no$-$evolution models.
This deviation is significant at 6.1$\sigma$ (3.7$\sigma$). 
A similar integration
is then performed in the same  redshift  range but at luminosities
L$_{0.5-2.0}>$1$\times10^{44}$erg s$^{-1}$. For the no$-$evolution model,
47.6 clusters are expected from the RASS1 (30.0 from the BCS) while
only 12 are observed, a factor 4 (2.5) less than predicted.  
Given the large uncertainties the significance of this result is 
at the 6.2$\sigma$ (3.9$\sigma$) level.
In agreement with the findings of \cite{vik98} and with the
recent results of the RDCS  (Rosati et al., this volume), 
we confirm a deficit of high luminosity high redshift clusters by a 
factor 2.5$-$4 which goes in the same  direction as the evolution of
the EMSS survey. However the deficit appears at
lower luminosities than in the EMSS ($>1\times10^{44}$ vs
$>3\times10^{44}$ in the 0.5$-$2.0 keV band), a result that we
are still investigating. At luminosities lower than $1\times10^{44}$ 
erg s$^{-1}$, no evidence for evolution is present, in agreement with 
the existing deep ROSAT cluster surveys.

\acknowledgements{Many people contributed to the construction and
success of the NEP survey. I would like to acknowledge the continuous
support and hard work of my colleagues at the IfA, Christopher Mullis 
and Patrick Henry. This work would not have been possible without the
collaboration of several MPE scientists and John Huchra. Partial 
financial support comes from NSF grant AST95$-$00515 and from CNR$-$ASI 
grants. Finally, I wish to thank Manolis Plionis, Ioannis Georgantopoulos 
and the other LOC members for organizing a great meeting in a 
magnificent Mediterranean spot.}

\vfill
\end{document}